\documentclass[lettersize,journal]{IEEEtran}
\usepackage{amsmath,amsfonts}
\usepackage{amsthm}
\usepackage{algorithmic}
\usepackage[ruled,linesnumbered]{algorithm2e}
\usepackage{array}
\usepackage[caption=false,font=normalsize,labelfont=sf,textfont=sf]{subfig}
\usepackage{textcomp}
\usepackage{stfloats}
\usepackage{url}
\usepackage{verbatim}
\usepackage{graphicx}
\usepackage{cite}
\usepackage{colortbl}  
\usepackage{xcolor}
\usepackage{url}

\usepackage{multirow}

\usepackage[colorlinks=true,
            linkcolor=blue]
           {hyperref}

\begin{document}

\title{Large Language Model-aided Edge Learning in Distribution System State Estimation}

\author{
        Renyou Xie,
        Xin Yin,
        Chaojie Li,
        Guo Chen,
        Nian Liu,
        Bo Zhao,
        Zhaoyang Dong
\thanks{
R. Xie, X. Yin, C. Li, G. Chen are with the School of Electrical Engineering and Telecommunications, the University of New South Wales, Sydney, 2052. E-mail: renyou.xie@unsw.edu.au;sinyinedu@163.com; chaojie.li@unsw.edu.au; guo.chen@unsw.edu.au.

 N. Liu, and B. Zhao are with the State Key Laboratory of Alternate Electrical Power System with Renewable Energy Sources, North China Electric Power University, Beijing, China. E-mail: nianliu@ncepu.edu.cn; zhaobozju@163.com).

Z. Y. Dong is with the Department of Electrical Engineering, City University of Hong Kong, Hong Kong. E-mail:zydong@ieee.org.
}}

\markboth{}%
{Shell \MakeLowercase{\textit{et al.}}: A Sample Article Using IEEEtran.cls for IEEE Journals}


\maketitle

\begin{abstract}
Distribution system state estimation (DSSE) plays a crucial role in the real-time monitoring, control, and operation of distribution networks. Besides intensive computational requirements, conventional DSSE methods need high-quality measurements to obtain accurate states, whereas missing values often occur due to sensor failures or communication delays.
To address these challenging issues, a forecast-then-estimate framework of edge learning is proposed for DSSE, leveraging large language models (LLMs) to forecast missing measurements and provide pseudo-measurements. 
Firstly, \textit{natural language-based prompts} and measurement sequences are integrated by the proposed LLM to learn patterns from historical data and provide accurate forecasting results. Secondly, a convolutional layer-based neural network model is introduced to improve the robustness of state estimation under missing measurement. Thirdly, to alleviate the overfitting of the deep learning-based DSSE, it is reformulated as a multi-task learning framework containing shared and task-specific layers. The uncertainty weighting algorithm is applied to find the optimal weights to balance different tasks. The numerical simulation on the Simbench case is used to demonstrate the effectiveness of the proposed forecast-then-estimate framework.
\end{abstract}


\section{Introduction}
The rapidly growing integration of distributed energy resources (DERs) and the increasing complexity of distribution networks have posed challenges to the control and operation of the distribution network.  DSSE can estimate the system state by utilizing the gathered measurements via IoT devices, which play a crucial role in the reliability and the stability of distribution networks~\cite{primadianto2016review,10173492}. However, due to the huge number of nodes and branches in the distribution network, it is incredibly difficult to solve it with high accuracy in real time under limited resources.

Typically, the algorithms of DSSE can be broadly classified into two categories: model-based approaches and data-driven approaches. The model-based DSSE method tries to map the measurement to the state by leveraging the accurate network topology and parameter information. The typical model-based algorithm for DSSE is weighted least squares~\cite{abur2004power}, which minimizes the sum of the squared differences between observed and predicted values. Although the weighted least square (WLS) approach is effective with redundant measurement, it is sensitive to outliers. To address this issue, the least absolute value (LAV) method is proposed by minimizing the absolute value of measurement. The LAV is more robust to WLS, whereas it is computationally expensive, and thus can hardly be applied to large-scale distribution networks~\cite{gol2014lav}. Except for these two methods, the extended Kalman filter (EKF) and the Uncented Kalman filter~\cite{huang2019robust, zhao2017robust} are adopted to implement DSSE considering the dynamic which can achieve a promising state estimation result. Nevertheless, these methods need to calculate the Jacobian matrices or Sigma points at each time step, which is computationally expensive and makes them hard to be applied in real-time.
In addition, the performance of the model-based method heavily depends on the accuracy of the network model and the availability of sufficient measurement data, which is hard to obtain in practice~\cite{bhela2017enhancing}. 

To overcome the limitations of model-based approaches, data-driven DSSE methods have gained increasing attention in recent years~\cite{9758816}. The data-driven methods utilize advanced machine learning or deep learning methods and abundant historical measurement data to learn a mapping function that maps measurement to states. Among various data-driven approaches, deep learning has shown promising results in improving the accuracy of DSSE~\cite{zhang_real-time_2019, 8586649}. 
 However, deep learning methods are prone to overfitting, especially when limited data are available. To improve the generalization performance of DSSE, Hybrid techniques such as ensemble learning~\cite{bhusal_deep_2021}, as well as the physical-informed neural networks (PINNs) are proposed~\cite{wu2022unrolled}.
In~\cite{zamzam_physics-aware_2020-1} the feed-forward neural network with graph-pruned structure mimics the physical topology of the distribution system, which forms a sparse weight matrix of the neural network.
Similarly,~\cite{pagnier_physics-informed_2021} uses the graph neural network that directly embeds the topology information to the estimation model to improve the DSSE performance.
Unlike the above method,~\cite{wang_physics-guided_2020, habib_deep_2023} integrate the power flow model into the neural network to improve the generalization performance. The state output by the neural network is fed into the physical model to get the estimated measurement, where the residual measurement is used to update the neural network. 
In~\cite{ngo2024physics}, the graph neural network is used to integrate the topology information, while a physical model is parallel implemented with estimated results to generate part of the loss function of the graph neural network model. 

Although the PINN is effective in alleviating the overfitting issue and making the data-driven part more interpretable, the training processing involves the complex power flow model, which increases the training complexity and slows down the training process. In addition, there are generally two tasks for state estimation, estimating the voltage magnitude and phase angle.
The aforementioned method either treats them as separate tasks or simply combines them as a single task, which ignores the correlation or difference between these tasks. This paper proposes to use the multi-task learning method to improve the generalization performance of the deep learning-based DSSE.

Another issue for the DSSE is the real-time data availability.
The increasing deployment of advanced metering infrastructure (AMI) and the emergence of edge computing and 5G  enables the processing of data closer to the source, reducing latency and improving privacy \cite{10024766}. Nevertheless,  the issue of missing or inaccurate measurements still remains due to sensor failures, and communication delays, which significantly deteriorate the performance of DSSE. To address this issue, the forecasting-based method is adopted. In \cite{bhusal_deep_2021}, the multivariate-linear regression is used to estimate the state for measurement imputation. In \cite{zhang_real-time_2019}, the RNN is adopted to forecast the state, where the forecasted state will be fed into the measurement function to generate a pseudo measurement, functioning as a replacement for the missing measurement. Lin et al.~\cite{lin_spatiotemporal_2024} argue that the connectivity of the power network results in the state interdependence between neighboring nodes. Therefore, a GCN-based method is proposed to fully leverage the spatial and temporal correlation between the power system state.
Despite the success of these forecasting-based methods, they still face challenges in capturing the complex dependencies and long-term temporal correlations in the measurement data. Moreover, these methods often require a substantial amount of historical data for training, which may not always be available in practice.
In recent years, large language models (LLMs) have emerged as a powerful tool for various natural language processing tasks, demonstrating remarkable performance in text generation, completion, and understanding~\cite{brown2020language, chowdhery2023palm, rae2021scaling}. These models, such as GPT-3.5~\cite{ouyang2022training}, GPT-4, and PaLM~\cite{chowdhery2023palm}, are pre-trained on vast amounts of text data, enabling them to learn rich representations and capture intricate patterns in the input sequences. Inspired by the success of these large language models, we propose a novel approach for DSSE that leverages its power to predict missing measurement values.

The contributions of the paper are as follows:

\begin{itemize}
    \item A LLM-based method is proposed to estimate the missing measurement to improve the state estimation performance. The proposed LLM-based forecasting method can seamlessly integrate language instructions and measurement sequences, enabling the model to learn valuable information from multiple data sources and provide highly accurate forecasting results.
    To the best of our knowledge, this is the first LLM that has been created for the problem of DSSE.
    \item  A deep learning-based method that combines convolutional layers and residual connections is proposed to implement distribution system state estimation. The proposed approach is robust and capable of handling DSSE tasks even in the presence of missing or imperfection measurements.
    \item The distribution system state estimation is reformulated as a multi-task learning problem for the first time for better generalization performance, where the uncertainty weighting algorithm is applied to find the optimal weight to balance different tasks.
\end{itemize}

\section{Distribution system state estimation problem formulation}
\subsection{Preliminary of DSSE}
\begin{figure}[h]
    \centering
\includegraphics[width=7cm,height=4.5cm]{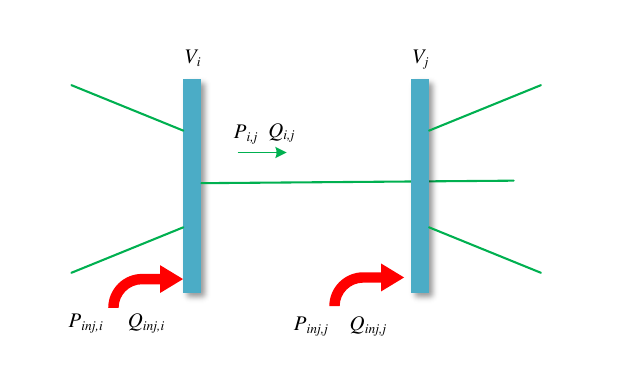}
    \caption{Power flow between buses.}
    \label{fig:power_flow}
\end{figure}

Briefly, the distribution system state estimation problem is to find the state $\boldsymbol{x}$ based on the measurement $\boldsymbol{z}$, where state $\boldsymbol{x}$ often refers to the voltage magnitude and angle, measurement $\boldsymbol{z}$ could be any type, such as active/reactive power injection, active/reactive lines power flow or bus voltage \cite{habib_deep_2023}, i.e.

\begin{subequations}
\begin{align}
    \boldsymbol{x} = [V_0,V_1,...,V_{n-1},  \theta_0=0,\theta_1,..,\theta_{n-1}] \in \mathbb{R}^{2n \times 1}, \\
    \boldsymbol{z} = [z_0,z_1,...,z_{m-1}] \in \mathbb{R}^{m \times 1},
\end{align}
\end{subequations}
where $V_i, \theta_i$,  represent the  voltage magnitude and phase angle of bus $i$ respectively; $n$, and $m$ are the number of buses and measurements. Specifically, there is a measurement function $h(\cdot)$ that maps $\boldsymbol{x}$ to $\boldsymbol{z}$. In the distribution system as depicted in \ref{fig:power_flow}, the measurement could be the different combination of the voltage magnitude or phase angle, active/reactive power injection, and active/reactive line power flows:

\begin{equation}
\left\{
\begin{subequations}
\begin{aligned}
    &P_{ij} = V_i^2 g_{ij} - V_iV_j(g_{ij}\cdot \cos(\theta_i - \theta_j) + b_{ij}\cdot \sin(\theta_i - \theta_j)) \\
    &Q_{ij} = -V_i ^2  b_{ij} - V_iV_j (g_{ij}\cdot \sin (\theta_i-\theta_j) - b_{ij}\cdot \cos(\theta_i - \theta_j)) \\
    &P_{inj,i} = \sum_{j\in \Omega_i} (P_{ji} - P_{ij})\\
    &Q_{inj,i} = \sum_{j\in \Omega_i} (Q_{ji} - Q_{ij})
\end{aligned}
\end{subequations}
\right.
\end{equation}
where $P_{ij}$, $Q_{ij}$ represent the active power and reactive power flow from bus $i$ to bus $j$ respectively; $g_{ij}$, $b_{ij}$ are the line conductance and susceptance, which are the real and imaginary part of the line admittance $Y_{ij}$ respectively; $P_{inj,i}, Q_{inj,i}$ are the active and reactive power injection of bus $i$. The first two subequations calculate the active and reactive power flowing from bus $i$ to bus
$j$, based on the voltage levels, phase angle differences and electrical properties of the transmission line. And the last two subequations determines the net active power injected at bus 
$i$ by summing the differences between incoming and outgoing active power flows. They are fundamental to power flow analysis, which determines the voltage, current, and power flows in a power system under a given load and generation condition.

\subsection{Problem Formulation}
There are lots of ways to find the $h^{-1}(\boldsymbol{z})$ to get the state $\boldsymbol{x}$ of the distribution systems.
These include model-based approaches such as the Weighted Least Squares (WLS) and the Least Absolute Value (LAV), as well as Bayesian model-based methods. Additionally, data-driven approaches such as deep neural networks and graph neural networks are also employed. Data plays a crucial role for both method types. The model-based methods typically require a measurement number 
 $m>2n$ to ensure observability, whereas data-driven methods need a substantial amount of data to develop a reliable model. This paper focuses on the data-driven approach due to the computational costs associated with model-based methods.

The conventional data-driven methods for DSSE can be expressed as 
\begin{align}
\min L(\boldsymbol{x},\boldsymbol{\hat{x}}), \notag \\
    s.t. \quad \boldsymbol{\hat{x}} = f(\bold{w};\boldsymbol{z}),
\end{align}
where the measurement $\boldsymbol{z}$ usually are $P_{ij}, Q_{ij}, P_{inj,i}, Q_{inj,i}$, and  $\bold{w}=[w_1, w_2,..,w_N]$ is the model parameter, and $\boldsymbol{\hat{x}}$ is the estimated stated obtained by the data-driven function $f$; The parameter $\bold{w}$ is updated by minimizing the loss $L(\boldsymbol{x},\boldsymbol{\hat{x}})$ with the gradient descent method.
There are two main problems existing in the data-driven-based methods: 1) the measurement $\boldsymbol{z}$ could contain missing values during the data collection and transmission procedure, which hurdle the state estimation process. For simplicity, the measurement that contains missing data is denoted as $\boldsymbol{\bar{z}}$; 2) the data-driven method is prone to overfitting, especially with limited data and training individual models for each state, i.e. ${\hat{x}_i} = f_i(\bold{w};\boldsymbol{z})$. This paper tries to find a more generalized model based on the imperfect measurement $\boldsymbol{\bar{z}}$, i.e.

\begin{align}
\min  \sum_i \lambda_i 
 L(\boldsymbol{x}_i,\boldsymbol{\hat{x}}_i), \notag \\
    s.t. \quad\boldsymbol{\hat{x}_i} = f(\bold{w};\boldsymbol{\bar{z}}),
\end{align}
where $\lambda_i$ is the weight of the corresponding loss $L$ for state $\boldsymbol{x}_i$, and the $\bold{w}=[w^{sh}, w^1, w^2,..,w^N]$, which involves the shared layers and task-specific layers. To solve the problem, this paper proposed a forecast-then-estimation framework by utilizing the large language model (LLM) and multitask learning (MTL).

\section{LLM-aided DSSE framework}
 In this section, the forecast-then-estimate framework for distribution system state estimation is presented in detail. The measurement forecasting using a large language model is first introduced. This is followed by a discussion of the multi-task learning method employed for training the station estimation model. Finally, the model used for state estimation is presented. 
 
\begin{figure}[h]
    \centering
\includegraphics[width=8cm,height=2cm]{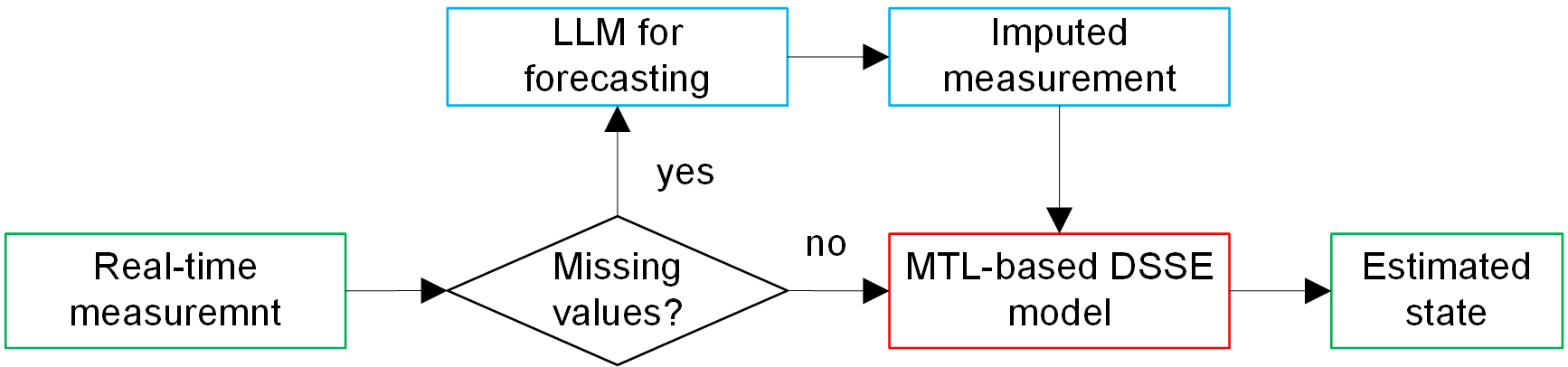}
    \caption{The forecast-then-estimate framework for real-time DSSE.}
    \label{fig:GPT_DSSE}
\end{figure}

 The proposed forecast-then-estimate framework is shown in Fig. \ref{fig:GPT_DSSE}. During the training stage, the forecasting model and state estimation model are trained separately. During the test stage, where real-time measurement is obtained, the framework will first decide whether there is a missing value in the measurement. If there is a missing value in the measurement, the LLM will forecast the corresponding measurement to act as a pseudo measurement and send it to the DSSE model to get the states. Otherwise, the DSSE model will directly use the real-time measurement for state estimation.

\subsection{ Measurement forecasting based on LLM}
\begin{figure}[h]
    \centering
\includegraphics[width=8.5cm,height=7cm]{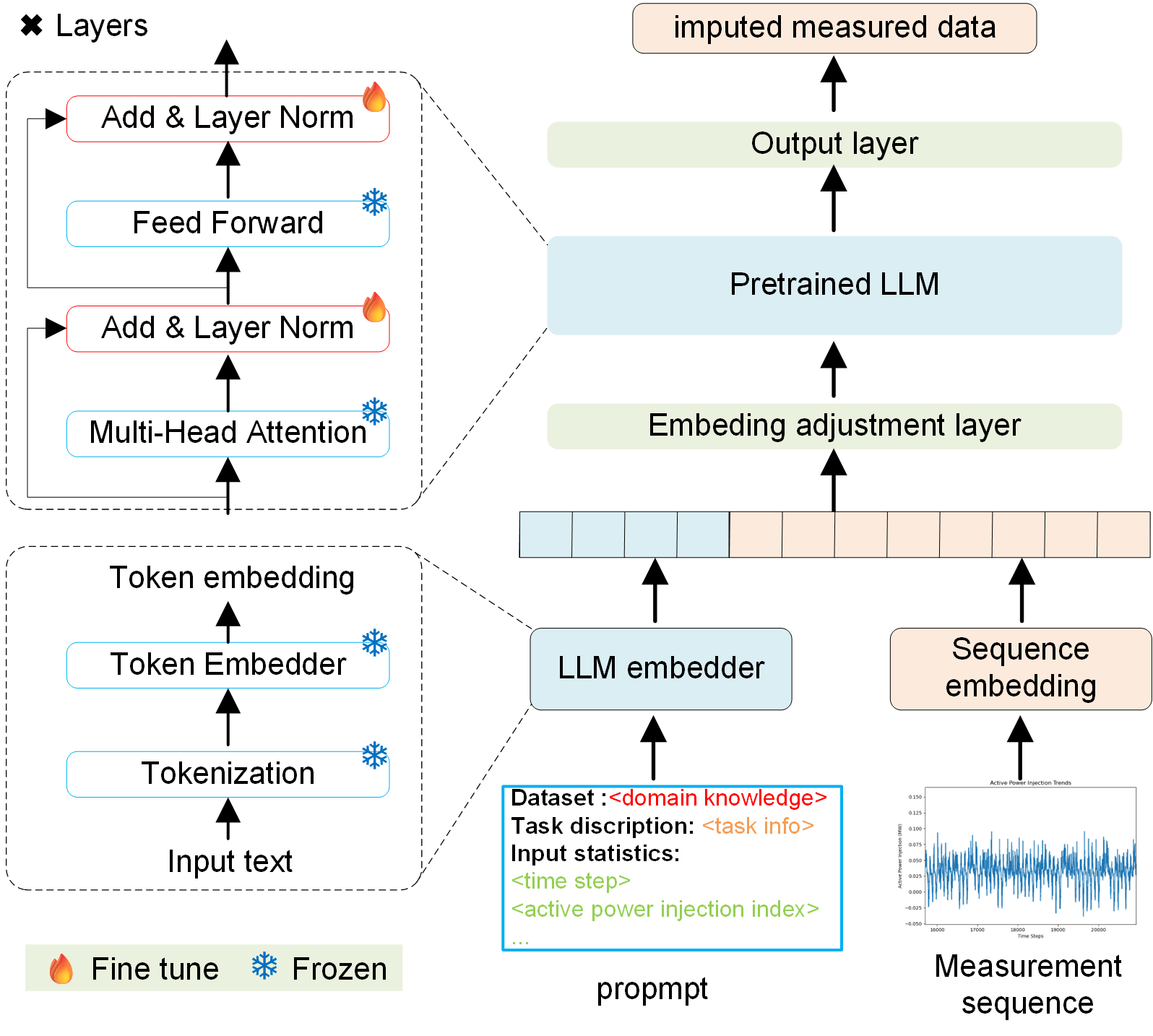}
    \caption{The measurement forecasting framework with LLM}
    \label{fig:GPT2}
\end{figure}

The data-driven DSSE assumes that all the measurements are completely available, whereas in real cases the measurements often contain the missing value during the data acquisition and transmission. Measurement forecasting is effective in addressing the problem of getting the pseudo measurement for state estimation. In the forecasting process, the former $k$ step of the measurement with missing value $[\boldsymbol{\bar{z}}(t-k),...,\boldsymbol{\bar{z}}(t)]$ is used to impute the missing value of $t$ step, i.e. to get $\boldsymbol{{z}}(t)$. Briefly, the forecasting task can be modeled as:
\begin{align}
   [\boldsymbol{\bar{z}}(t-k),...,\boldsymbol{\bar{z}}(t)] \xrightarrow{\mathcal{F}(\cdot)} \boldsymbol{{z}}(t),
\end{align}
where $\mathcal{F}(\cdot)$ is the model used for forecasting. In this paper, the large language model is adopted as the forecasting model to impute the missing measurement in real time. The diagram of using the LLM for missing measurement forecasting is shown in Fig. \ref{fig:GPT2}.

\subsubsection{Distribution system measurement forecasting process} 

As indicated in Fig.\ref{fig:GPT2}, the basic process of utilizing the LLM for measurement forecast is as follows:

\begin{itemize}
    \item Embedding. Given the input measurement data $\boldsymbol{\bar{z}} = [\boldsymbol{\bar{z}}(t-k),...,\boldsymbol{\bar{z}}(t)]$ and the corresponding language prompt $\boldsymbol{p}$, the embedding layer converts them into continuous vector representations $\mathbf{E}_z \in \mathbb{R}^{L_s \times D}$ and $\mathbf{E}_p \in \mathbb{R}^{L_p \times D}$, respectively, where $L_s$ is the sequence length of the measurement data, $L_p$ is the sequence length of the prompt, and $D$ is the embedding dimension. The embedding layer can be represented as:
    \begin{equation}
    \mathbf{E}_z = \text{Embedding}_z(\boldsymbol{\bar{z}}) + \mathbf{P}_z,
    \end{equation}
    \begin{equation}
    \mathbf{E}_p = \text{Embedding}_p(\boldsymbol{p}) + \mathbf{P}_p,
    \end{equation}
    where $\text{Embedding}_z(\cdot)$ and $\text{Embedding}_p(\cdot)$ are the embedding functions for the measurement data and prompt, respectively, and $\mathbf{P}_z \in \mathbb{R}^{L_s \times D}$ and $\mathbf{P}_p \in \mathbb{R}^{L_p \times D}$ are the position encodings used to capture the temporal information.
    Note that the prompt is an optional choice for measurement forecasting.
    \item Prompt concatenation. The embedded measurement sequence $\mathbf{E}_z$ and prompt $\mathbf{E}_p$ are concatenated to form a single input sequence $\mathbf{E} \in \mathbb{R}^{(L_s+L_p) \times D}$:
\begin{equation}
\mathbf{E} = [\mathbf{E}_p; \mathbf{E}_z].
\end{equation}

    \item Embedding adjustment. The concatenated input $\mathbf{E}$ is passed through an embedding adjustment layer to adapt and specialize the representation for improved forecasting performance:
\begin{equation}
\mathbf{E}_{adj} = \text{SeqEmbedding}(\mathbf{E}).
\end{equation}

\item LLM processing. The adjusted embedding $\mathbf{E}$ is fed into the pre-trained LLM, which generates a contextualized representation $\mathbf{H} \in \mathbb{R}^D$ capturing the semantic relationships between the prompt and measurement data:
\begin{equation}
\mathbf{H} = \text{LLM}(\mathbf{E}_{adj}).
\end{equation}

\item  Output generation. Finally, the contextualized representation $\mathbf{H}$ is passed through an output layer to generate the predicted measurement sequence $\hat{\mathbf{z}} \in \mathbb{R}^{L_o}$, where $L_o$ is the length of the output sequence:
\begin{equation}
\hat{\mathbf{z}} = \text{OutputLayer}(\mathbf{H}).
\end{equation}
\end{itemize}

\subsubsection{GPT-2 Architecture}

In this paper, the GPT-2 \cite{radford2019language} is selected as the pre-trained LLM for distribution system measurement forecast.
The GPT-2 model utilizes the Transformer decoder architecture and consists of multiple stacked decoder blocks. Each decoder block includes two sub-layers: multi-head self-attention mechanism and feed-forward neural network, which is depicted in Fig. \ref{fig:GPT2}.

The multi-head self-attention mechanism is used to capture the long-range dependencies in the measurement data and the prompt. For the $i$-th head, the self-attention mechanism can be represented as \cite{vaswani2017attention}:
\begin{equation}
\text{head}_i = \text{Attention}(\mathbf{Q}_i, \mathbf{K}_i, \mathbf{V}_i) = \text{softmax}(\frac{\mathbf{Q}_i\mathbf{K}_i^T}{\sqrt{d_k}})\mathbf{V}_i,
\end{equation}
where $\mathbf{Q}_i, \mathbf{K}_i, \mathbf{V}_i \in \mathbb{R}^{L \times d_k}$ are the query, key, and value matrices, respectively, $d_k=D/h_{att}$ is the dimension of each head, and $h_{att}$ is the number of heads. The multi-head self-attention mechanism concatenates the outputs of all heads and applies a linear transformation:

\begin{equation}
\text{MultiHead}(\mathbf{Q}, \mathbf{K}, \mathbf{V}) = \text{Concat}(\text{head}_1, ..., \text{head}_h)\mathbf{W}^O,
\end{equation}
where $\mathbf{W}^O \in \mathbb{R}^{D \times D}$ is the linear transformation matrix.

Feed-Forward Neural Network: The feed-forward neural network applies a non-linear transformation to the output of the self-attention mechanism, enhancing the model's expressiveness. Moreover, to facilitate gradient propagation and model training stability, residual connections and layer normalization are applied after each sub-layer:
\begin{equation}
\mathbf{x} + \text{Sublayer}(\text{LayerNorm}(\mathbf{x})),
\end{equation}
where $\text{Sublayer}(\cdot)$ represents the self-attention mechanism or the feed-forward neural network sub-layer. In this paper, the layer norms are set as the tunable parameter during the training process.

Overall, the GPT-2-based measurement forecasting process offers a robust and efficient solution for addressing the challenge of missing measurements in power system state estimation.

\subsection{Multi-task learning for DSSE}
\begin{figure}[h]
    \centering
\includegraphics[width=9cm,height=4.2cm]{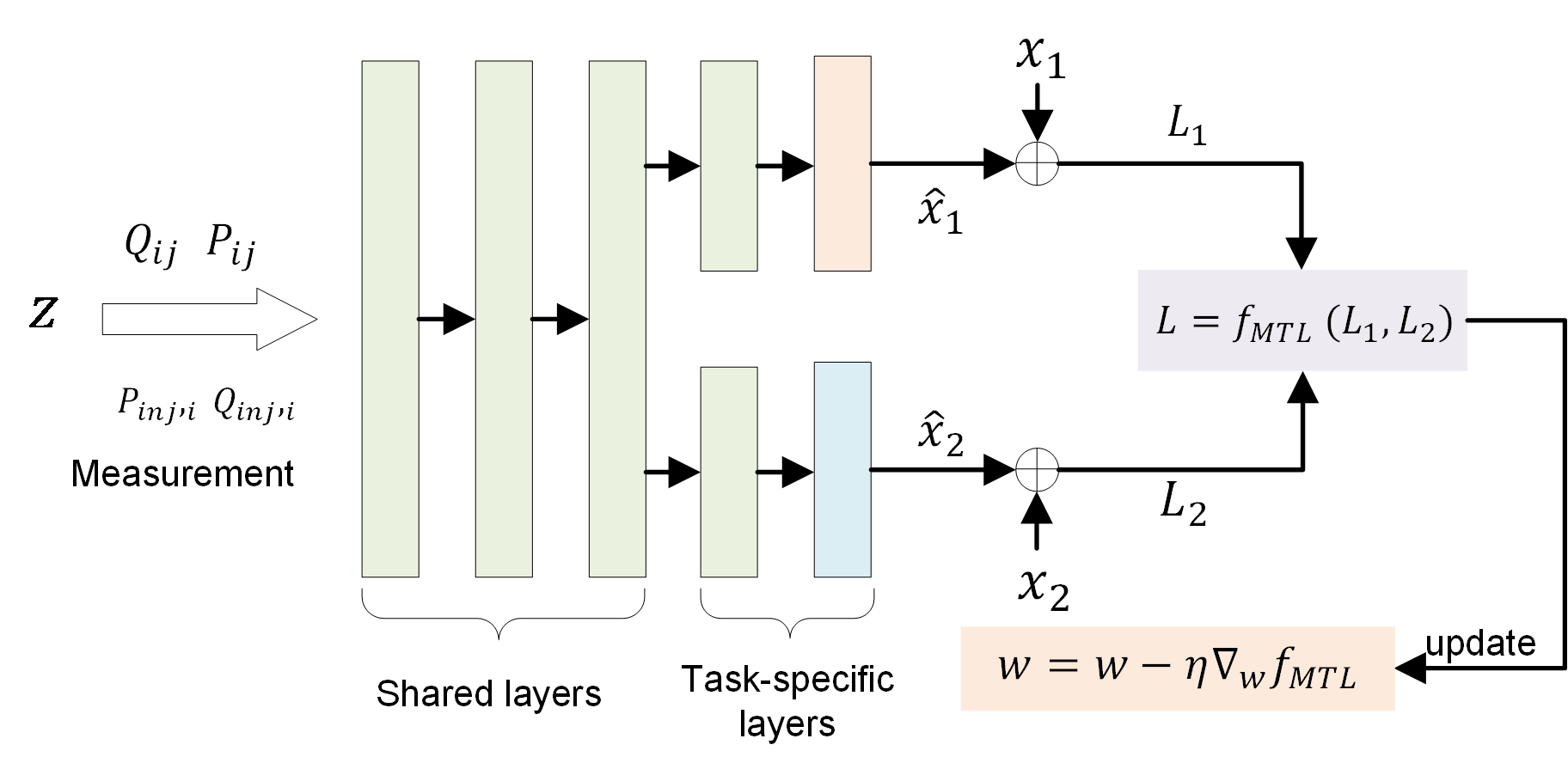}
    \caption{Multi-task learning framework for DSSE}
    \label{fig:MTL2}
\end{figure}

In distribution system state estimation, the voltage magnitude and phase angle of each bus node are treated as the state. The naive way for data-driven DSSE is to build two models for the magnitude and angle estimation separately. However, the single-task learning methods ignore the highly related relationship between the voltage magnitude and phase angle, causing the overfitting issue. To address this issue, we leverage the multitask learning technique and build a single model to implement these two tasks by attributing the voltage magnitude and angle estimation task-specific layers.  For better demonstration, the multitask learning framework is shown in Fig. \ref{fig:MTL2}

As indicated in Fig. \ref{fig:MTL2}, the proposed structure is composed of 2 parts: the shared layers, and the task-specific layers. The shared layer takes the measurement $\boldsymbol{z}$ as the input and learns the shared structure between the state $\boldsymbol{x}_1$  (voltage magnitude) and state $\boldsymbol{x}_2$ (voltage angle); After that, the output of the shared layer will be feed into the task-specific layers to get the estimated state $\boldsymbol{\hat{x}}_1$, $\boldsymbol{\hat{x}}_2$. The estimated states are compared with the true state to get the loss $L_1(\boldsymbol{x}_1, \boldsymbol{\hat{x}}_1)$, $L_2(\boldsymbol{x}_2, \boldsymbol{\hat{x}}_2)$ for parameter updating.  To effectively update the whole model, the three loss functions are combined to get the final loss function: 
\begin{align}
    L = f_{MTL}(L_1,L_2),
\end{align}
 where $f_{MTL}$ is a function that combines the two loss functions. The commonly used method is  weighted loss,  i.e $f_{MTL}(L_1, L_2)=\sum_{i=1}^2 \lambda_i L_i$, where $\lambda_i$ is the weight attribute to the corresponding loss. Nevertheless, the optimal weight is hard to obtain. In this paper, the $f_{MTL}$ is determined by the uncertainty weighting algorithm (UWA) proposed in \cite{Kendall_2018_CVPR}, which can dynamically optimize the weights attributed to different losses. The basic idea of the uncertainty weighting method is to maximize the Gaussian likelihood with homoscedastic uncertainty. Assume $f(\bold{w};\boldsymbol{z})$ is the output of the neural network with parameter $\bold{w}$, and the input $\boldsymbol{z}$, the likelihood of the desired output $\bold{x}$ is modeled as 
 \begin{align}
     p(\boldsymbol{x}|f(\bold{w};\boldsymbol{z})) = \mathcal{N}(f(\bold{w};\boldsymbol{z}), \sigma^2),
 \end{align}
where $\sigma$ represents the observation noise scale. When there are multiple tasks, the multi-task likelihood can be modeled as :
\begin{align}
    p(\boldsymbol{x}_1,..., \boldsymbol{x}_N |  f(\bold{w};\boldsymbol{z}))
=p(\boldsymbol{x}_1|f(\bold{w};\boldsymbol{z}))...p(\boldsymbol{x}_N|f(\bold{w};\boldsymbol{z})),
\end{align}
 where $N$ is the task number. In our case, $N=2$, and $\boldsymbol{x}_1$, $\boldsymbol{x}_2$ represent the voltage magnitude, voltage phase angle respectively. The state estimation aims to maximize the likelihood $p(\boldsymbol{x}_1,\boldsymbol{x}_2 |  f(\bold{w};\boldsymbol{z}))$, which means to maximize the log-likelihood of the model, i.e.
 \begin{align}\label{eqn:log_likelihood}
     &\max_{\bold{w},\sigma} \log p(\boldsymbol{x}_1,\boldsymbol{x}_2 |  f(\bold{w};\boldsymbol{z})) \notag \\ 
     &= \log p(\boldsymbol{x}_1|  f(\bold{w};\boldsymbol{z})) + \log p(\boldsymbol{x}_2| f(\bold{w};\boldsymbol{z})) \notag \\
     &= \mathcal{N}(\boldsymbol{x}_1;f(\bold{w};\boldsymbol{z}), \sigma_1^2) + \mathcal{N}(\boldsymbol{x}_2;f(\bold{w};\boldsymbol{z}), \sigma_1^2),
 \end{align}
 since the state estimation task can be considered a regression task, the log-likelihood can be rewritten as 
 \begin{align}\label{eqn:log_likelihood2}
\log p(\boldsymbol{x}|f(\bold{w};\boldsymbol{z}))\propto -\frac{1}{2\sigma^2}||f(\bold{w};\boldsymbol{z})-\boldsymbol{x}||^2 - \log \sigma.
 \end{align}
Therefore  \eqref{eqn:log_likelihood} can be written as :
\begin{align}
    &\min_{\bold{w},\sigma} -\log p(\boldsymbol{x}_1,\boldsymbol{x}_2|  f(\bold{w};\boldsymbol{z})) \notag \notag \\
     \propto & \frac{1}{2\sigma_1^2}||f(\bold{w};\boldsymbol{z})-\boldsymbol{x}_1||^2 +  \frac{1}{2\sigma_2^2}||f(\bold{w};\boldsymbol{z})-\boldsymbol{x}_2||^2 +
     \log \sigma_1 \sigma_2 \notag \\
    =&     \frac{1}{2\sigma_1^2} L_1(\bold{w}) + \frac{1}{2\sigma_2^2} L_2(\bold{w}) + \log \sigma_1 \sigma_2, 
\end{align}
where $L_1(\bold{w})=||f(\bold{w};\boldsymbol{z})-\boldsymbol{x}_1||^2$ is the loss function for voltage magnitude $\boldsymbol{x}_1$, similarly applies for $L_2(\bold{w})$. Note that $L_1(\bold{w})$, $L_2(\bold{w})$ are equivalent to 
$L_1(\boldsymbol{x}, \boldsymbol{\hat{x}})$, $L_2(\boldsymbol{x}_2, \boldsymbol{\hat{x}_2})$ respectively. Besides, $\sigma_1$, and $\sigma_2$ are trainable parameters that represent the noise level of the corresponding output and balance the weight attributed to different tasks. For example, when $\sigma_1$ increases, the weight attributed to $L_1(\bold{w})$ will decrease. Moreover, to avoid the three loss term being minimized by setting high $\sigma_1, \sigma_2$, the $\log \sigma_1 \sigma_2$ is added as the penalized term.

\subsection{Model structure}
This subsection introduces the model structure that is used to estimate the state of the distribution system.

\begin{figure*}[h]
    \centering
    \includegraphics[width=14cm,height=3cm]{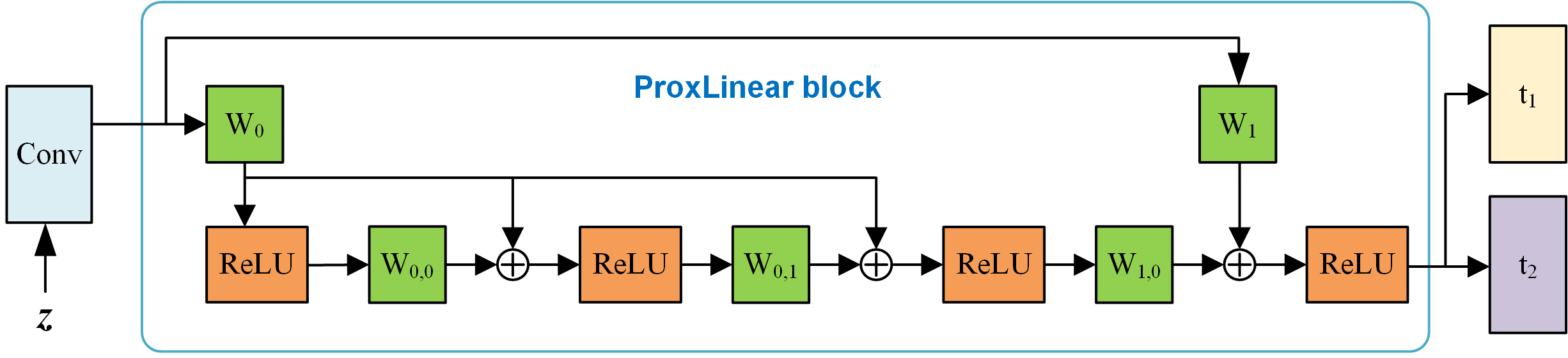}
    \caption{The CNN-Prox model structure for DSSE}
    \label{fig:CNN_Prox}
\end{figure*}

The multi-task learning improves the generalization performance by designing the training framework and the optimized weighting strategy for different task losses. To further improve the generalization of the neural network model, the residual net connection idea is used in this paper. There are two effective residual connection-based models for state estimation, the ProxLinear Net \cite{zhang_real-time_2019}, and ResNetD \cite{bhusal_deep_2021}. 
In this paper, the ProxLinear module is selected as the foundational component due to its straightforward architecture and capability to emulate the solver used in model-based methods. Additionally, a convolutional layer is incorporated to enhance the ability of the ProxLinear Net to handle measurements with missing values. The model employed in this study is abbreviated as CNN-Prox, and its structure is depicted in Fig. \ref{fig:CNN_Prox}. The "Conv“ represents the convolution layer, $t_1$ represents the task-specific layers for the voltage magnitude estimation task, while $t_2$ is the layers for the voltage phase angle estimation task.

\section{Cases study}
To testify the efficacy of the proposed method, case studies based on the Simbench \cite{en13123290} database are implemented. 
Specifically,  a medium voltage distribution network named "1-MV-urban--0-sw" is used to generate the state estimation data, in which Pandapower \cite{thurner2018pandapower} are used. 
The network "1-MV-urban--0-sw" contains 144 nodes, where 139 nodes are loaded and 134 are equipped with generation units including Wind power, PV, and Hydropower. The data contains a one-year load profile with a revolution of 15 minutes, where the load of one month is selected to generate the measurement and states for state estimation verification. The active/reactive power injection and active/reactive line power flows are considered as the measurement, while the voltage magnitude and angle are considered as states. In this scenario, 60\% of the nodes are assumed to place measurement devices.

To get a generalized model, the dataset is split into training, validation, and test sets, which share 50$\%$, 10$\%$, and 40$\%$ of the whole dataset. During the training phase, the model with minimal loss on the validation set is selected as the final model for testing to alleviate the overfitting issue. For the forecasting task,  the Adam optimizer is adopted with a learning rate of 0.0001, where the mean squared loss is used as the criterion to update the model parameter. The sequence length is set to 96 (corresponding to 24h) to get the prediction. Similarly, for the state estimation task, the Adam optimizer is used with a learning rate of 0.001.  Huber loss is used as the loss function to alleviate the influence of abnormal values. 

To evaluate the performance of the proposed method comprehensively, the  Mean absolute error (MAE), and Root mean square error (RMSE) are used as the metrics, i.e.
\begin{align}
    MAE = \frac{1}{N} \sum_{i=1}^N|y_i - \hat{y}_i|,\\
    RMSE = \sqrt{\frac{\sum_{i}^N (y_i-\hat{y}_i)^2}{N}},
\end{align} 
where $y_i$ represents the true value of $i$-th measurement or state, and the $\hat{y}_i$ is the estimated state or measurement.

To demonstrate the efficacy of the proposed framework, we conducted a series of comparisons. First, we compared the forecasting results of GPT-2 with those of conventional methods. Subsequently, we evaluated the performance of the state estimation model CNN Prox against conventional methods, while also comparing the multi-task learning algorithm. Finally, we showcased the performance of the forecasting-then-estimation framework by applying it to real measurements containing missing values.

\subsection{Forecasting performance of GPT-2}
\begin{figure}[h]
    \centering
    \includegraphics[width=8.5cm,height=4cm]{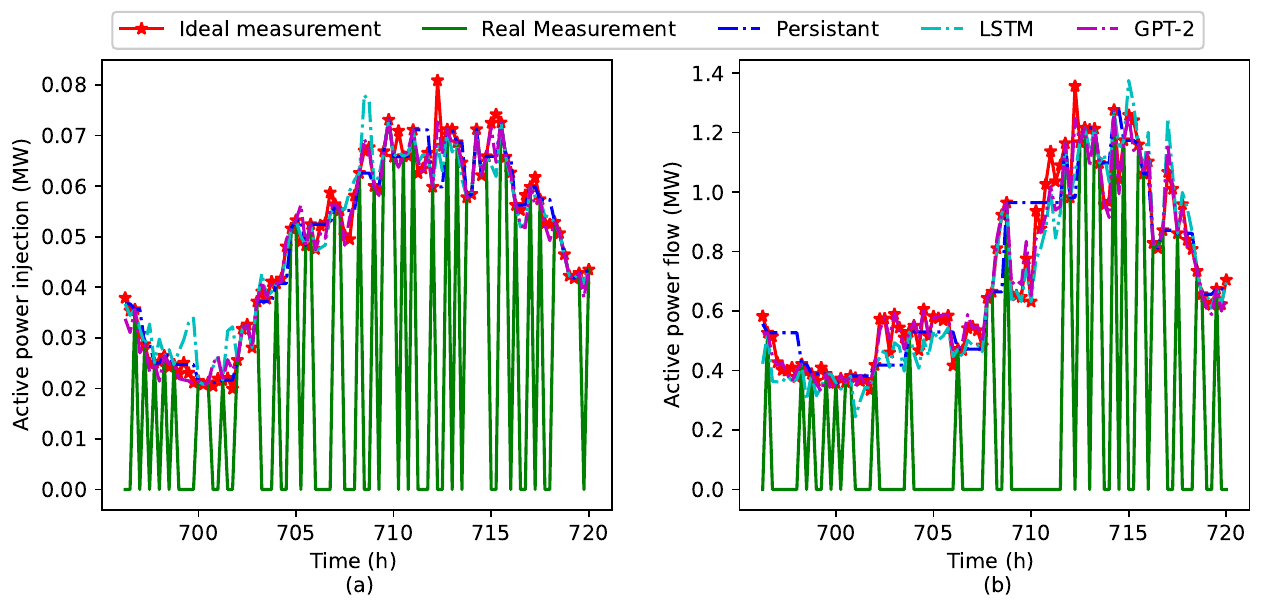}
    \caption{Measurement forecasting results of different methods on (a) power injection (b) power flow }
    \label{fig:forecast_compare}
\end{figure}

In real-world scenarios, measurements from distribution systems often contain missing values, which can hinder the state estimation process. To address this issue, the proposed approach utilizes a Large Language Model (LLM), specifically GPT-2, to forecast or impute the missing values. This approach takes advantage of the excellent performance of LLMs in handling language and their potential for time series forecasting. To demonstrate the performance of the proposed measurement forecasting framework using GPT-2, it is compared with two typical forecasting methods: Long Short-Term Memory (LSTM) and a persistent model. Note that for the persistence model, the last non-missing value is used to fill the missing value of the current step. For simulating real measurements with missing values, a predefined percentage (called missing ratio) of the ideal measurements is selected and set as "NaN". The missing ratio is set to [0.1, 0.2, 0.3, 0.4, 0.5] to evaluate the performance of GPT-2. The results of this comparison are presented in Table \ref{table:forecasting_compare}. As indicated in Table \ref{table:forecasting_compare}, the GPT-2 $>$ LSTM $>$ persistence model both on MAE and RMSE metrics under various, where $">"$ represents better than. For better demonstration, the measurement (active power injection and active power flow) of a node and line in the network is shown in Fig. \ref{fig:forecast_compare}. The missing ratio of the real measurement is set to 0.5 for demonstration, whereas the "NaN" in real measurement is set to 0. As indicated in Fig. \ref{fig:forecast_compare}, the real measurements differ much from the ideal measurement, The persistence model provides a naive way to fill the missing values, but it fails to capture the dynamic change of the measurement, especially when there are consecutive missing values. The LSTM can predict the dynamic change of the measurement, whereas the predicted values could differ much from the measurement. The GPT-2 can fit the measurement well and provide more accurate forecasting results. The RMSE of persistence, LSTM, and GPT-2 are 0.1058, 0.05661, and 0.04855 respectively.

 \begin{table}[h]
	\renewcommand{\arraystretch}{1.1}
	\caption{Performance of different forecasting methods under various missing ratios}
	\centering
	\label{table:forecasting_compare}
	\centering

	\resizebox{\linewidth}{!}{
		\begin{tabular}{c  l c  c   c  c c}
			\hline\hline \\[-3mm]
			 \multirow{2}{*}{Metrics} & \multirow{2}{*}{Methods} & \multicolumn{5}{c}{Missing ratio} 
			\\ 
			\cline{3-7}
			&   &0.1 &0.2 &0.3 &0.4 &0.5 \\
			\hline
             \multirow{4}{*}{MAE} &persistence & 0.02751&0.03014 &0.03216 &0.03390 &0.03911\\
             &LSTM &0.01179 &0.01388 &0.01524 &0.01775 &0.02093\\
             &GPT-2 &0.007589 &0.01116 &0.01311 &0.01553 &0.01872\\

           \hline
            \multirow{4}{*}{RMSE} &persistence &0.06994 &0.08473 &0.09264 &0.08911 &0.1058\\
             &LSTM &0.02883 &0.03735 &0.04304 &0.04844 &0.05661\\
              &GPT-2 &0.02059 &0.02915 &0.03399 &0.03904 &0.04855\\

			\hline\hline
		\end{tabular}
		}
\end{table}

\subsection{Influence of the multi-task learning}
\begin{figure}[h]
    \centering
    \includegraphics[width=9cm,height=6cm]{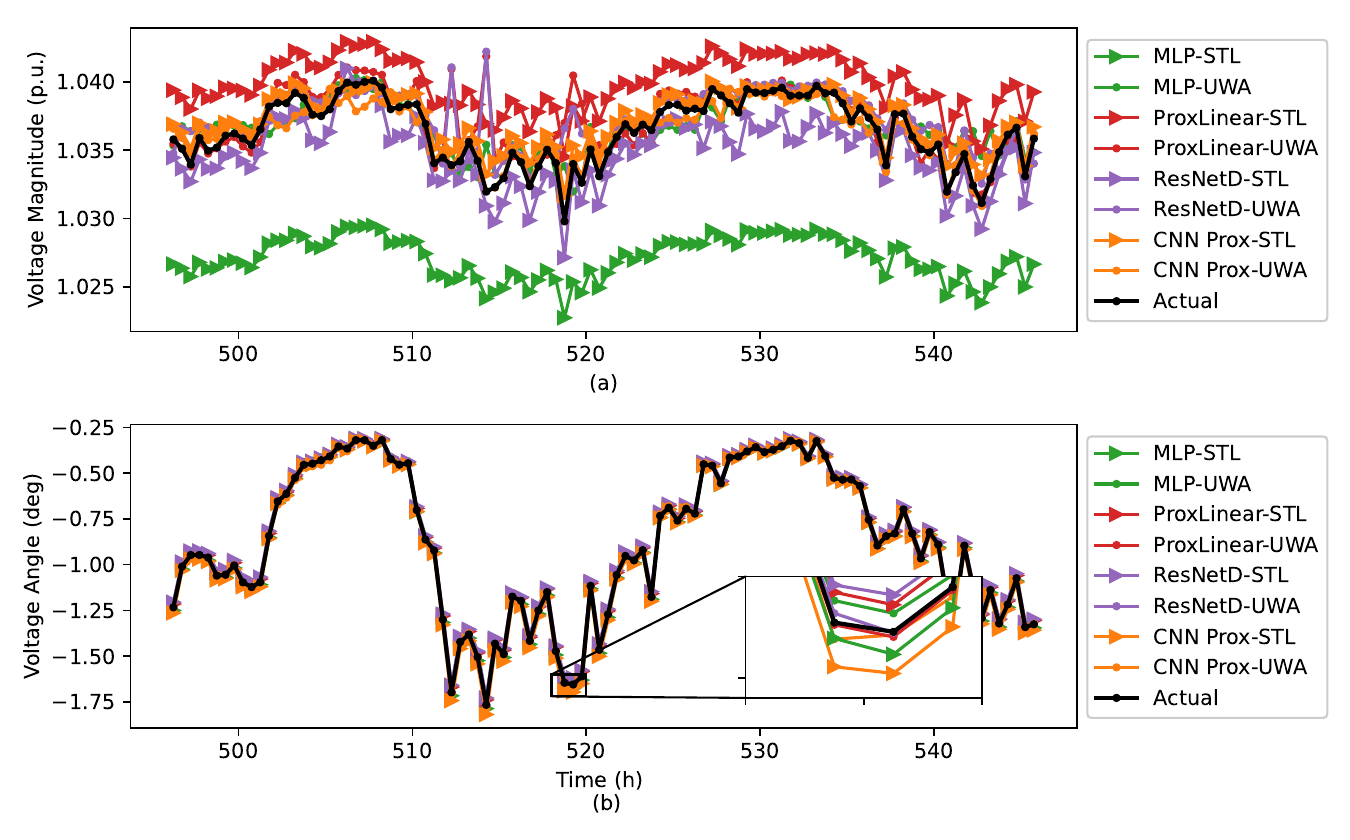}
    \caption{State estimation results of different models and learning framework on (a) voltage magnitude (b) voltage angle.}
    \label{fig:DSSE_model_MTL_compare}
\end{figure}

Conventional state estimation treats the Voltage magnitude and voltage angle as separated tasks and trains the model for each task, which ignores the correlation of each and is prone to overfitting. This paper adopts the multi-task learning framework to address this issue by using the uncertainty weighting algorithm (UWA). To demonstrate the effectiveness of the UWA, the single-task learning (STL) method with the same neural network structure, where the voltage magnitude or angle is considered as a single task. Specifically, the commonly used method Multi-layer perception (MLP), and the ProxLinear Net, ResNetD, as well as the proposed CNN-Prox are used to testify to the improvement of the UWA, and the Results are shown in Table \ref{table:DSSE_compare}, and  Fig. \ref{fig:DSSE_model_MTL_compare}. As indicated in Table \ref{table:DSSE_compare}, the MAE and RMSE for all the methods drop after applying the multi-task learning and UWA, except that the ResNetD on voltage magnitude estimation task, demonstrating the effectiveness of the UWA. For the Voltage magnitude estimation task, the MAE of all methods drops over 30\%,  with MLP's MAE dropping from 0.00957 to 0.0011 after applying UWA. The improved performance of UWA on the voltage magnitude task can be easily observed in Fig. \ref{fig:DSSE_model_MTL_compare} (a), where the STL versions of MLP, ProxLinear Net, and ResNetD fail to fit the true values.
For the voltage angle estimation task, the MAE drops over 50\% for all models. Among the four methods, the ProxLinear Net archives the best performance for voltage angle
estimation, while the CNN-Prox gets the minimal MAE and RMSE under both STL and UWA for voltage estimation tasks.

\begin{table}[h]
	\renewcommand{\arraystretch}{1.1}
	\caption{DSSE results of different models under single task and multi-task learning framework}
	\centering
	\label{table:DSSE_compare}
	\centering

	\resizebox{1\linewidth}{!}{
		\begin{tabular}{c  l c  c  p{0.1cm} c  c }
			\hline\hline \\[-3mm]
			 \multirow{2}{*}{Metrics} & \multirow{2}{*}{Model} & \multicolumn{2}{c}{Voltage Magnitude} & &\multicolumn{2}{c}{Voltage Angle} 
			\\ 
			\cline{3-4} \cline{6-7}
			&   &STL  &UWA  & &STL  &UWA\\
			\hline
             \multirow{3}{*}{MAE} &MLP &0.00957 &\textbf{0.00111} &  &0.01369  &\textbf{0.00729}\\
             &ProxLinear Net &0.00346  &\textbf{0.00089}  & &0.01147  &\textbf{0.00364} \\
             &ResNetD &0.00145 &\textbf{0.00109}  & &0.01342 &\textbf{0.00492}\\
            &CNN-Prox &0.00090  &\textbf{0.00066} & &0.01373  &\textbf{0.00643}\\

           \hline
             \multirow{3}{*}{RMSE} &MLP & 0.00962 &\textbf{0.00143} & &0.01569  &\textbf{0.00999}\\
             &ProxLinear Net &0.00351  &\textbf{0.00158} & &0.01325  &\textbf{0.00504}\\
             &ResNetD &\textbf{0.00161}  &0.00188 & &0.01699  &\textbf{0.00644}\\
            &CNN-Prox  &0.00119  &\textbf{0.00086} & &0.01621 &\textbf{0.00894}\\

			\hline\hline
		\end{tabular}
		}
\end{table}

\begin{figure}[h]
    \centering
    \includegraphics[width=8.5cm,height=4cm]{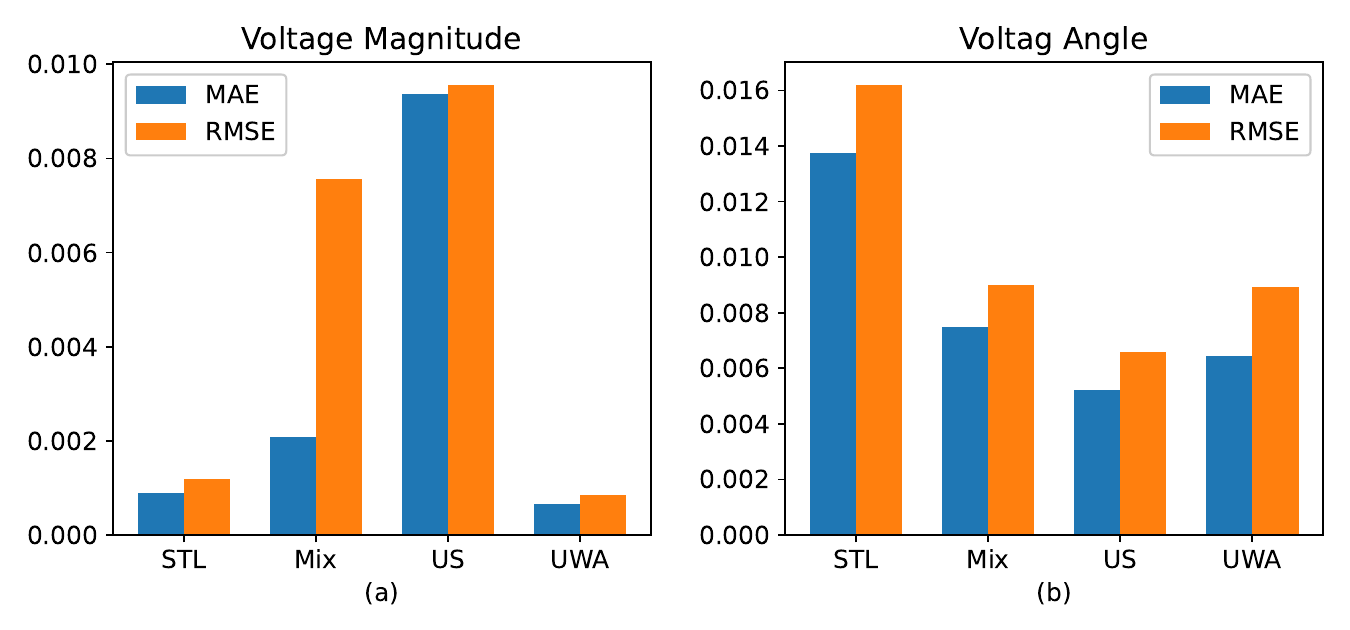}
    \caption{State estimation result of CNN-Prox with various learning methods on (a) voltage magnitude (b) voltage angle. }
    \label{fig:DSSE_MTL_compare_bar}
\end{figure}

To further testify to the performance of UWA, it is compared with the STL, uniform scaling (US) method, and mixture (Mix) of tasks, where the tasks share all the layers. For demonstration, only the results for CNN-Prox are listed in Table \ref{table:compare_MTL}, with its bar diagram shown in Fig. \ref{fig:DSSE_MTL_compare_bar}. For voltage magnitude estimation task, both US and Mix cannot even outperform the STL, whereas the UWA gets the best performance. For the voltage angle task, both US, Mix, and UWA outperform STL, and US achieves the lowest MAE and RMSE, 0.00522 and 0.00657 respectively. Although US outperforms UWA on voltage angle tasks with slightly lower MAE and RMSE values, its MAE and RMSE on voltage magnitude are 10 times higher than that of UWA. Consequently, the UWA is a preferred MTL framework for DSSE.

\begin{table}[h]
	\renewcommand{\arraystretch}{1.1}
	\caption{DSSE results for CNN-Prox with various learning frameworks}
	\centering
	\label{table:compare_MTL}
	\centering

	\resizebox{1\linewidth}{!}{
		\begin{tabular}{c  c c  c   c  c }
			\hline\hline \\[-3mm]
			 \multirow{2}{*}{States} & \multirow{2}{*}{Metrics} & \multicolumn{4}{c}{Methods} 
			\\ 
			\cline{3-6}
			&   &STL &Mix &US &UWA \\
			\hline
             \multirow{2}{*}{Magnitude}  
            &MAE &0.00090 &0.00208 &0.00936 &0.00066\\
            &RMSE &0.00119 &0.00757 &0.00957 &0.00086\\

           \hline
             \multirow{2}{*}{Angle}  
        &MAE &0.01373 &0.00750 &0.00522 &0.00643\\
        &RMSE  &0.01621 &0.00899 &0.00657 &0.00894\\
    
			\hline\hline
		\end{tabular}
		}
\end{table}

\subsection{Performance of forecasting-aided state estimation }
To testify the effectiveness of the GPT model for measurement forecasting, the forecasted measurements are used as inputs to the state estimation models, such as CNN-Prox, MLP, ProxLinear, and ResNetD. The estimated states are then compared with those obtained using other forecasting methods, including the persistence method and LSTM, as well as the ideal and real measurements.

\begin{figure}[h]
    \centering
    \includegraphics[width=9cm,height=6cm]{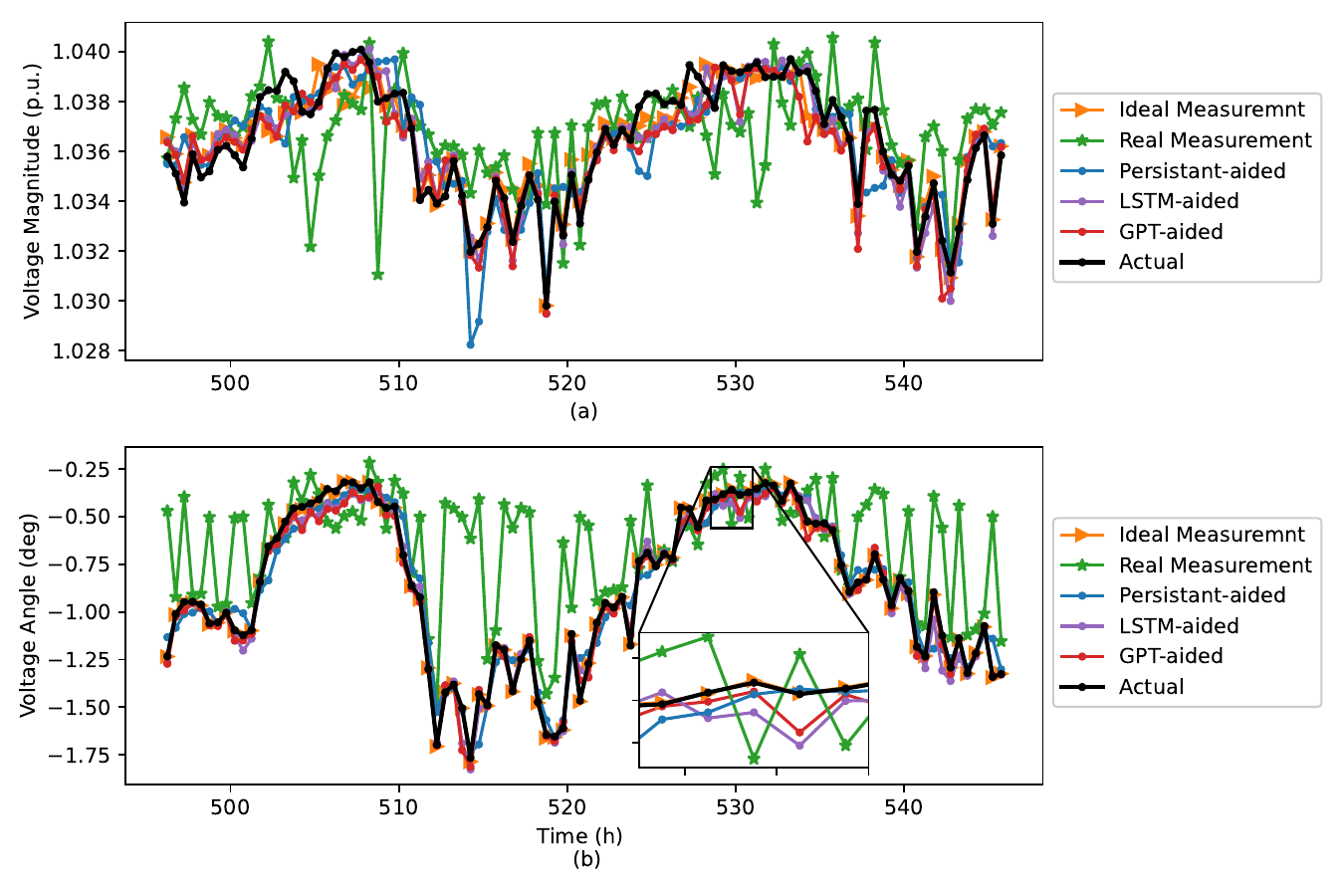}
    \caption{State estimation result of CNN-Prox with various measurements on (a) voltage magnitude, (b) voltage phase angle}
    \label{fig:DSSEforecast_compare}
\end{figure}

Fig. \ref{fig:DSSEforecast_compare} presents the estimated voltage magnitude and angle curves from 510h to 540h using different measurement sources. It can be observed that there exists a noticeable gap between the estimated states using ideal and real measurements, while the forecasting methods can help the state estimation models approximate the actual values to varying degrees. Among them, using GPT-2 to forecast the input measurements demonstrates better performance in estimating the actual states, especially for voltage angles.

Table \ref{table:compare_with_others_response_R1} quantifies the estimation errors of different models using various measurement sources. When utilizing the GPT-2-aided measurements as inputs, the estimation errors of all models are generally lower compared to those using the persistent and LSTM methods. For the CNN-Prox model, the MAE and RMSE of voltage magnitude estimation using GPT-2-aided measurements are 0.00068  and 0.00089, respectively, which are 24.4\% and 25.8\% lower than those using persistent-aided measurements. The improvement is more prominent in voltage angle estimation, where the MAE and RMSE are reduced by 55.0\% (from 0.05556  to 0.02504) and 57.1\% (from 0.08485 to 0.03635), respectively, compared to the persistence-aided case. These results indicate that incorporating GPT-2 for measurement forecasting can effectively enhance the accuracy of distribution system state estimation.

Among the estimation models, the CNN-Prox exhibits superior performance with relatively smaller errors when using the same input measurements for voltage magnitude estimation. With GPT-2-aided measurements, the CNN-Prox model achieves an MAE of 0.00068 p.u. and an RMSE of 0.00089 p.u. for voltage magnitude estimation, which are 36.4\% and 35.0\% lower than those of the MLP model, respectively. However, for voltage angle estimation, the CNN-Prox model's performance is not as good as ProxLinear and ResNetD when using GPT-2 or LSTM-aided measurements. The inferior performance of CNN-Prox on Voltage angle estimation is consistent with the result using the ideal measurement as the input (discussed in the former part of this section). 
Interestingly, when the input data differs significantly from the ideal measurement, such as in the case of real measurements or the persistence-aided measurement, the CNN-Prox model outperforms other methods in voltage angle estimation. For instance, with real measurements, the CNN-Prox model achieves an MAE of 0.21896  and an RMSE of 0.32381  for voltage angle estimation, which is 51.6$\%$ and 38.2$\%$ lower than those of the ResNetD model (0.45270  and 0.52393), respectively. This finding demonstrates that the CNN-Prox model is more robust to missing values and data imperfections, particularly in the context of voltage angle estimation.

However, it is worth noting that ideal measurements are usually unavailable in practice. By introducing measurement forecasting techniques, especially advanced models like GPT-2, the issues related to low-quality real measurements can be mitigated, providing a promising direction for further improving the distribution system state estimation. 

 \begin{table}[h]
	\renewcommand{\arraystretch}{1.1}
	\caption{Voltage magnitude estimation result with different data source}
	\centering
	\label{table:compare_with_others_response_R1}
	\centering

	\resizebox{\linewidth}{!}{
		\begin{tabular}{p{0.7cm}  l c  c   c  c }
			\hline\hline \\[-3mm]
			 \multirow{2}{*}{Metrics} & \multirow{2}{*}{Data Source} & \multicolumn{4}{c}{Models} 
			\\ 
			\cline{3-6}
			&   &MLP &ProxLinear & ResNetD & CNN-Prox \\
			\hline
           \hline
            \multirow{5}{*}{MAE} &Ideal measurement &0.00111 &0.00089 &0.00109 &0.00066\\
            &Real measurement  &0.01267 &0.00752 &0.00402  & 0.00169\\
            &persistence-aided&0.00115 &0.00099 &0.00114 &0.00090\\
             &LSTM-aided &0.00107 &0.00086 &0.00107  &0.00068\\
             &GPT-2-aided &0.00107 &0.00086 &0.00103  &0.00068\\

           \hline
            \multirow{5}{*}{RMSE} &Ideal measurement &0.00143 &0.00158 &0.00188 &0.00086\\
            &Real measurement &0.01518 &0.00888 &0.00501 &0.00230\\
            &persistence-aided&0.00147 &0.00137 & 0.00158 &0.00120\\
             &LSTM-aided &0.00137 &0.00138 &0.00163  &0.00089\\
             &GPT-2-aided &0.00137 &0.00143 &0.00138  &0.00089\\

			\hline\hline
		\end{tabular}
		}
\end{table}


 \begin{table}[h]
	\renewcommand{\arraystretch}{1.1}
	\caption{Voltage Angle estimation result with different data source}
	\centering
	\label{table:compare_with_others_response_R1}
	\centering

	\resizebox{\linewidth}{!}{
		\begin{tabular}{p{0.7cm}  l c  c   c  c }
			\hline\hline \\[-3mm]
			 \multirow{2}{*}{Metrics} & \multirow{2}{*}{Data Source} & \multicolumn{4}{c}{Models} 
			\\ 
			\cline{3-6}
			&   &MLP &ProxLinear & ResNetD & CNN-Prox \\
			\hline
           \hline
            \multirow{5}{*}{MAE} &Ideal measurement &0.00729 &0.00364 &0.00492 &0.00643 \\
            &Real measurement  &0.46626 &0.46071 &0.4527 &0.21896 \\
            &persistence-aided &0.06384 &0.06304 &0.06048 &0.05556 \\
             &LSTM-aided &0.02902 &0.02808 &0.02767 &0.02814\\
             &GPT-2-aided  &0.02491 &0.02347 & 0.02329 &0.02504 \\

           \hline
            \multirow{5}{*}{RMSE} &Ideal measurement &0.00999 &0.00504 &0.00644 &0.00894\\
            &Real measurement &0.53664 &0.53566 &0.52393 &0.32381\\
            &persistence-aided&0.08886 &0.08833 &0.08543  &0.08485\\
             &LSTM-aided &0.04081 &0.04000 &0.03951  &0.04169\\
             &GPT-2-aided & 0.03446 &0.03317 &0.03299 &0.03635\\

			\hline\hline
		\end{tabular}
		}
\end{table}

\begin{figure}[h]
    \centering
    \includegraphics[width=8.5cm,height=4cm]{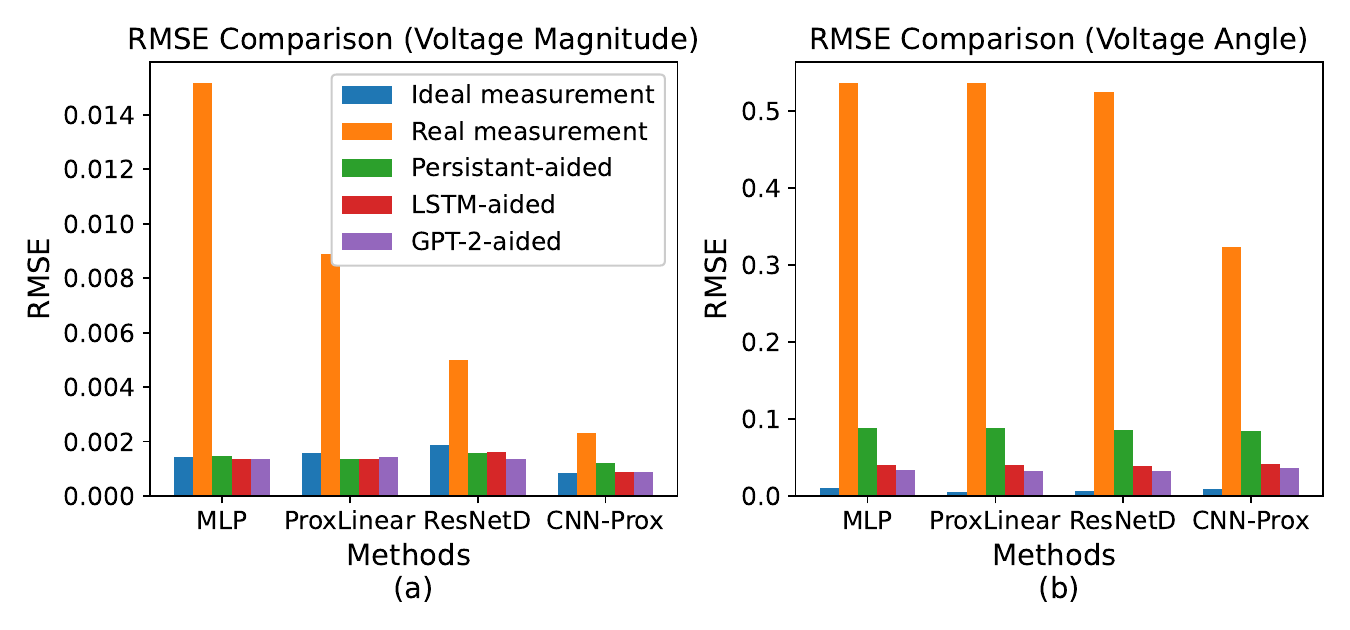}
    \caption{ RMSE of various models with different measurements on (a) voltage magnitude, (b) voltage phase angle.   }
    \label{fig:DSSE_forecast}
\end{figure}

\section{Discussion}
Although the proposed method can better handle the voltage magnitude and angle estimation tasks, and make an accurate state estimation with missing measurement, there are still several limitations.

First of all, the missing value prediction is implemented using the large language model, which could be cumbersome and require a lot of memory to be deployed, especially when using a complex model such as LLaMa 3.1, with a parameter of 405 billion. Therefore, the model quantization technique could be a potential future work for better application of large language models with limited hardware resources.

Secondly, only the missing values are considered in the real measurement, whereas there could be anomaly values or false data in the measurement. Therefore, integrating the anomaly detection function in the forecasting model could be more beneficial. A possible way is to consider the missing value imputing and anomaly detection a multitask learning problem, and use the uncertainty weight loss for training a comprehensive model to deal with missing and abnormal value simultaneously.

In addition, although the multitask learning improves the state estimation performance by finding the common features as well as the difference between tasks, it is still a data-driven approach. The black-box nature makes it hard to explain. Therefore, integrating physical information could benefit the data-driven method and provide more reliable results.

\section{Conclusion}
In this paper, a forecast-then-estimate framework is proposed to ensure robust distribution system state estimation with missing measurements. A large language model-based approach is first proposed to forecast the missing measurements in real time to provide pseudo-measurements for state estimation. The proposed LLM model can integrate language instructions and measurement sequences, enabling the model to learn valuable information and provide accurate forecasting results.
Moreover, the voltage magnitude and phase angle estimation tasks are formulated as a multi-task learning framework, where the uncertainty weighting algorithm (UWA) is adopted to obtain optimal weights. Finally, CNN-Prox is proposed to be used as the state estimation model to achieve better estimation results with missing measurements. The effectiveness of the proposed framework is demonstrated on the Simbench case, with comparisons to conventional methods. Results indicate that the proposed LLM-based forecasting method can accurately forecast the missing values, and multi-task learning with UWA can improve the performance of both tasks compared to single-task learning methods. In addition, the proposed CNN-Prox outperforms conventional methods for the voltage magnitude estimation task while achieving competitive performance on the voltage phase angle estimation task.

\bibliographystyle{IEEEtran}
\bibliography{reference}

\end{document}